\documentclass[a4paper,11pt]{article}
\usepackage{jheppub} 

\usepackage{physics}
\usepackage{longtable}
\usepackage{float}
\usepackage{amsthm}
\usepackage{amssymb}
\usepackage{hyperref}
\numberwithin{equation}{section}
\usepackage{graphicx}%
\usepackage{multirow}%
\usepackage{amsmath,amssymb,amsfonts}%
\usepackage{amsthm}%
\usepackage{mathrsfs}%
\usepackage[title]{appendix}%
\usepackage{xcolor}%
\usepackage{textcomp}%
\usepackage{manyfoot}%
\usepackage{booktabs}%
\usepackage{algorithm2e}
\usepackage{algpseudocode}%
\usepackage{listings}%

\theoremstyle{thmstyleone}%
\theoremstyle{thmstyletwo}%
\newtheorem{remark}{Remark}%

\newtheorem{conjecture}{Conjecture}%
\theoremstyle{thmstylethree}%
\newtheorem{definition}{Definition}%

\def\C{\mathbb{C}}
\def\Z{\mathbb{Z}}
\def\Q{\mathbb{Q}}
\def\R{\mathbb{R}}

\def\P{\mathbb{P}}

\def\Hirz[#1]{\mathbbm{F}_{#1}}
\def\o[#1]{\overline{#1}}

\def\prim{\text{prim}}

\def\ker{\mbox{ker}}

\def\C{\mathbb{C}}
\def\Z{\mathbb{Z}}
\def\Q{\mathbb{Q}}
\def\R{\mathbb{R}}

\def\P{\mathbb{P}}

\def\Hirz[#1]{\mathbbm{F}_{#1}}
\def\o[#1]{\overline{#1}}

\def\prim{\text{prim}}

\def\ker{\mbox{ker}}

\author{Hugo Fortin, }
\author{Daniel Lopez Garcia}

\emailAdd{hugo@fortin.io}
\emailAdd{danielflg@id.uff.br}

\title{Hodge theory and $G_4$ fluxes in weighted projective spaces: Galois action}

\abstract{
We extend the explicit study of \(G_4\)-fluxes and general Hodge cycles from the ordinary Fermat sextic fourfold to tame Fermat-type hypersurfaces in weighted projective space.  The main new feature in the weighted setting is that the Galois action on the cyclotomic period data need not preserve the \((2,2)\)-subspace.  As a consequence, the rational reconstruction of an integral self-dual class can involve additional middle-cohomology components, increasing the norm of the corresponding flux.

We work at maximally symmetric Fermat points, where the period matrices and symmetry-invariant Hodge loci can be computed explicitly.  Using Movasati's description of periods, cyclotomic period matrices, and Hermite/Smith normal form reductions, we construct the relevant integral lattices of symmetric self-dual classes in middle cohomology.  This gives a controlled test of whether symmetric general Hodge cycles can satisfy the M2-brane tadpole bound.

Our main conclusion is empirical.  In the degree 12 example in \(\mathbb{P}_{1,1,1,1,4,4}\), and in the degree 8 weighted example used as a comparison, the shortest computed symmetric general Hodge cycles overshoot the tadpole bound.  In the degree 36 example in \(\mathbb{P}_{1,1,1,9,12,12}\), which has \(h^{1,1}=11\), the most general example we have, the tadpole conjecture is indeed verified at the maximally symmetric locus, although the computations get difficult and computationally expensive.

These computations suggest that, in weighted Fermat examples, the relevant notion of a ``symmetric flux'' must take into account not only automorphisms of the variety but also the Galois action on the period field.  Non-uniform Galois orbits provide a natural arithmetic mechanism by which symmetric Hodge classes can acquire large tadpole charge.
}

\begin{document}

\maketitle

\section{Introduction}

The tadpole conjecture, first made in \cite{Bena_2021} and studied algorithmically in \cite{bena2021algorithmicallysolvingtadpoleproblem}, implies that there is a tension between the number of stabilized moduli and the tadpole bound, potentially isolating vacua from the vast number thought of previously \cite{Denef_2004}\cite{Taylor_2015}. 
Many studies have appeared in relation with the tadpole conjecture, such as \cite{Plauschinn_2022},\cite{Grimm_2022} or, similarly to the following albeit focusing on AdS vacua : \cite{lüst2022tadpoleconjectureinteriormoduli}. All of them find evidence towards the tadpole conjecture.

Thus, it is natural to ask further characterization of the tadpole conjecture in mathematical terms, as well as ways to avoid it. One such way was found in \cite{Becker:2022LGtadpoles}, with notable success \cite{Rajaguru:2024emw}, or alternatively in \cite{chen2025symmetriesmtheorylikevacuadimensions} for symmetry-based arguments. Furthermore, one can argue that symmetric fluxes lead to small tadpoles \cite{Coudarchet:2023mmm}. Lastly, one can consider singular geometry by arguing that the tadpole conjecture holds only for smooth Calabi-Yau varieties, and find the associated gauge algebras as was done in \cite{braun2023tadpolesgaugesymmetries}.

The main motivation for the present work is to follow up the work on \cite{braun2024gfluxgeneralhodgecycles}, assuming \cite{Braun_2021}, by extending the results to Minkowski vacua associated with more general Fermat-type hypersurfaces and to offer further characterization on what "symmetric fluxes" mean. 

The present work should be viewed as a continuation of the analysis of general Hodge cycles on the Fermat sextic.  Our goal is not to prove a global non-existence theorem for flux vacua in weighted projective hypersurfaces.  Rather, we ask what happens to the flux lattice, the tadpole bound, and the arithmetic of the period matrix once one moves away from the ordinary projective case.

The main lesson is that weighted Fermat hypersurfaces introduce a new arithmetic effect.  The Galois orbit of a residue form of type \((2,2)\) need not remain inside \(H^{2,2}\oplus H^{2,2}\).  When the Galois completion of the period data is used to reconstruct integral classes, this non-uniformity can enlarge the resulting norm.  This gives a concrete mechanism by which even symmetric Hodge classes may become expensive with respect to the tadpole bound.

Our evidence is necessarily sectorial.  We work at maximally symmetric Fermat points and compute the corresponding symmetry-invariant lattices. In all examples computed, we found that the tadpole conjecture is verified and that the shortest primitive general Hodge cycle is above the tadpole bound.

\subsection{Flux vacua and the Gukov--Vafa--Witten superpotential}

The Hodge-theoretic quantities studied in this paper have a direct physical interpretation in the four-dimensional $\mathcal{N}=1$ effective supergravity obtained from M-/F-theory compactified on a Calabi--Yau fourfold $X$. The key ingredient is the Gukov--Vafa--Witten (GVW) flux superpotential \cite{Gukov:1999ya,Dasgupta:1999ss,Haack_2001}.

In the M-theory description, a background flux
\[
   G \in H^4(X,\R)
\]
satisfying the quantization condition \eqref{eq:flux-quantization} and the supersymmetry conditions \eqref{eq:primitive-hodge} induces a superpotential
\begin{equation}
   W_G
   = \int_{X} \Omega \wedge G ,
   \label{eq:GVW-superpotential}
\end{equation}
where $\Omega$ is a nowhere-vanishing holomorphic $(4,0)$-form. This is the natural analogue, for Calabi--Yau fourfolds, of the type IIB superpotential
\[
   W_{G_3}(\tau,z)
   = \int_{Y} G_3 \wedge \Omega_3,
\]
with $G_3 = F_3 - \tau H_3$ on a Calabi--Yau threefold $Y$.

\medskip

To make the link with Hodge theory explicit, fix a basis $\{\gamma_I\}$ of $H_4(X,\Z)$ and let $\{\omega^I\}$ be the dual basis of $H^4(X,\Z)$, so that
\[
   \int_{\gamma_I}\omega^J = \delta_I^{\ J}.
\]
We define the period vector of $\Omega$ by
\[
   \Pi^I := \int_{\gamma_I} \Omega,
\]
and expand the flux as
\[
   G = \sum_I N_I \,\omega^I,
   \qquad
   N_I := \int_{\gamma_I} G \in \Z \left(+\tfrac{1}{2}\Z\right),
\]
where the half-integer shift is determined by the parity of $c_2(X)$ in \eqref{eq:flux-quantization}. The superpotential \eqref{eq:GVW-superpotential} can then be written as
\begin{equation}
   W_G = \sum_I N_I \,\Pi^I,
   \label{eq:GVW-periods}
\end{equation}
so it is linear in the flux quanta $N_I$, with coefficients given by periods of $\Omega$. All the dependence on complex structure is therefore encoded in the variation of Hodge structure on $H^4(X)$.

\medskip

The complex structure moduli $T^a$ are chiral multiplets in the four-dimensional $\mathcal{N}=1$ theory, with Kähler potential
\[
   K_{\rm cs}
   = -\log\Bigl( i^{\,4} \int_{X_T} \Omega \wedge \overline{\Omega} \Bigr)
\]
and superpotential $W_G(T)$ as above. The F-term equations for the complex structure are
\begin{equation}
   D_a W_G := \partial_a W_G + (\partial_a K_{\rm cs}) W_G = 0,
   \label{eq:Fterm-complex}
\end{equation}
where $\partial_a$ is over the complex structure moduli. Griffiths transversality gives
\[
   \partial_a \Omega \in H^{3,1}(X),
\]
and hence
\[
   \partial_a W_G(T)
   = \int_{X} \partial_a \Omega \wedge G.
\]
Thus the F-term equations \eqref{eq:Fterm-complex} constrain the $(3,1)$ and $(1,3)$ components of $G$ and, at a supersymmetric point, force the flux to be of Hodge type $(2,2)$ and primitive. This is exactly the condition that $G$ defines a Hodge class in $H^{2,2}(X)$, as in the mathematical discussion below.

\medskip

For the Kähler moduli we assume, as is standard in flux compactifications, that the flux does not generate a tree-level superpotential; their stabilization requires additional ingredients (non-perturbative effects) and will not play a role here. Throughout the paper we therefore keep the Kähler moduli fixed and focus solely on the complex structure sector, with the understanding that physically relevant fluxes are those whose $(2,2)$ components are also primitive, i.e.\ satisfy $G\wedge\omega=0$.

\medskip

The scalar potential restricted to complex structure moduli is
\[
   V_{\rm cs}
   = e^{K_{\rm cs}}\Bigl(
        K^{a\bar b}_{\rm cs}\, D_a W_G\,\overline{D_b W_G}
        - 3 |W_G|^2
      \Bigr),
\]
where $K^{a\bar b}_{\rm cs}$ is the inverse Kähler metric on moduli space. A supersymmetric Minkowski vacuum in this sector requires
\begin{equation}
   W_G = 0,
   \qquad
   D_a W_G = 0 \quad \forall a.
   \label{eq:Minkowski-cond}
\end{equation}
We do not study the branch \(W_G\neq 0\) here, and werestrict to fluxes supporting supersymmetric Minkowski vacua. For subtleties
in interpreting nonzero-flux-superpotential backgrounds as controlled static
string/M-theory vacua, see \cite{Sethi:2017phn}. In what follows we are interested in fluxes that can support Minkowski vacua, i.e.\ classes $G$ for which \eqref{eq:Minkowski-cond} has a solution.

\medskip

Expanding $W_G$ around a candidate vacuum $T_0$ we have
\[
   W_G = W_G(T_0)
          + \frac{1}{2}\sum_{a,b} H_{ab} (T^a-T_0^a)(T^b-T_0^b)
          + \cdots,
\]
with Hessian
\[
   H_{ab} = \left.\frac{\partial^2 W_G}{\partial T^a \partial T^b}\right|_{T_0}.
\]
The Hessian controls the quadratic fluctuations of the potential. We say that a flux $G$ defines a general Hodge cycle if, at the corresponding Hodge locus, the matrix $H_{ab}$ has maximal rank in the complex-structure directions: physically, this means that all complex structure moduli are stabilized at the level of the Hessian, and the remaining massless fields span the kernel of $H_{ab}$.

\section{Mathematical preliminaries}

Let $X_T$ be a family of Calabi--Yau fourfolds parametrized by points $T$ in a complex algebraic base
\[
   B \ni T \longmapsto X_T .
\]
We denote by 
\[
   H^4_T := H^4(X_T,\mathbb{Z})/\mathrm{torsion}
\]
the middle cohomology, endowed with its natural polarized integral variation of Hodge structure of weight~$4$. We recall the Hodge decomposition
\begin{align*}
    H^4_T \otimes \mathbb{C} 
      &= \bigoplus_{p+q=4} H^{p,q}(X_T), \\
    H^{p,q}(X_T) &= \overline{H^{q,p}(X_T)} ,
\end{align*}
and we write $h^{p,q}(X_T) = \dim H^{p,q}(X_T)$ for the Hodge numbers. The Euler characteristic of a fiber will be denoted by $\chi(X_T)$, or simply $\chi(X)$ when no confusion can arise.

\medskip

Let $\omega_T \in H^{1,1}(X_T)\cap H^2(X_T,\mathbb{R})$ be the Kähler class and let 
\[
  L_T : H^k(X_T) \longrightarrow H^{k+2}(X_T),\qquad
  L_T(\alpha) = \omega_T \wedge \alpha
\]
be the Lefschetz operator. We will mostly work with the primitive part of cohomology:

\begin{definition}
  A class $\alpha \in H^k(X_T)$ is called primitive if 
  \[
     L_T^{4-k+1}(\alpha) = 0 .
  \]
  We denote by $H^k_{\mathrm{prim}}(X_T)$ the corresponding primitive subspace.
\end{definition}

On the primitive middle cohomology $H^4_{\mathrm{prim}}(X_T)$ there is a polarization
\[
   Q_T : H^4_{\mathrm{prim}}(X_T) \times H^4_{\mathrm{prim}}(X_T) \longrightarrow \mathbb{Z},
\]
which is symmetric and non-degenerate. We also consider the associated Weil operator
\[
   C_T : H^4_T \otimes \mathbb{R} \longrightarrow H^4_T \otimes \mathbb{R},
\]
acting on $H^{p,q}(X_T)$ by $C_T|_{H^{p,q}} = i^{\,p-q}$. A class $v \in H^4_T \otimes \mathbb{R}$ is called self-dual (in the sense of \cite{BakkerGrimmSchnellTsimerman:2023}) if
\[
   C_T v = v,
\]
and any Hodge class of type $(2,2)$ is automatically self-dual in this sense.

\subsection{Hodge classes, Hodge loci and general Hodge cycles}

We denote by
\[
   H^{2,2}(X_T)_{\mathbb{Q}} := H^{2,2}(X_T)\cap H^4(X_T,\mathbb{Q})
\]
the space of rational Hodge classes of type $(2,2)$, and similarly by
\[
   H^{2,2}(X_T)_{\mathbb{Z}} := H^{2,2}(X_T)\cap H^4(X_T,\mathbb{Z})
\]
the lattice of integral Hodge classes. Given a local section
\[
   G \in H^4_{\mathrm{prim}}(X_T)\cap H^4_T
\]
of the local system, we define its Hodge locus:

\begin{definition}
  For a fixed class $G \in H^{2,2}({T_0})$, the Hodge locus (or Noether--Lefschetz locus) of $G$ is
  \[
     V_G = \{\, T \in B \mid G_T \in H^{2,2}(X_T) \,\},
  \]
  where $G_T$ denotes the parallel transport of $G$ along the variation of Hodge structure.
\end{definition}

The union of all such loci
\[
   \mathrm{Hdg}(X) := \bigcup_{G \in H^4_{T_0}} V_G
\]
is the Hodge locus in $B$.

A cycle $\delta\in H^4(X_T, \mathbb{Q})$ satisfying the nullity conditions in the periods with $p\neq q$ is known as a \textbf{Hodge cycle}. 

A particularly important role will be played by Hodge classes that stabilize all complex structure moduli in the physical applications. In the language of \cite{braun2024gfluxgeneralhodgecycles} these correspond to general Hodge cycles: for such a class $G$, the matrix of second derivatives of the superpotential has maximal rank and therefore no complex structure moduli remain unstabilized.

In what follows, contrary to what was worked out in \cite{braun2024gfluxgeneralhodgecycles}, we do not assume the Hodge conjecture: instead we work solely with Griffiths residues, and make no statement on the status of the Hodge conjecture for the examples we study, nor do we compute algebraic cycles and their intersections. It would nonetheless be interesting to study the (computational) Hodge conjecture in those examples.

\subsection{Fluxes, quantization and the tadpole bound}

In the M-/F-theory setting, a flux configuration on a Calabi--Yau fourfold is encoded by a cohomology class
\[
   G \in H^4(X,\mathbb{R})
\]
satisfying the quantization condition
\begin{equation}
   G + \frac{1}{2} c_2(X) \in H^4(X,\mathbb{Z})
   \label{eq:flux-quantization}
\end{equation}
and the supersymmetry conditions
\begin{equation}
   G \in H^{2,2}(X), \qquad G \wedge \omega = 0.
   \label{eq:primitive-hodge}
\end{equation}
The quantization condition leads naturally to integer number of $M_2$ branes \cite{Klemm_1998}.
Thus supersymmetric Minkowski vacua correspond to primitive Hodge classes in the shifted lattice
\[
   \Lambda_{\mathrm{phys}} 
     = \Bigl\{\, G \in H^{2,2}(X)_{\mathbb{R}} \cap H^4_{\mathrm{prim}}(X)
                  \ \Big|\ G + \tfrac{1}{2}c_2(X) \in H^4(X,\mathbb{Z}) \Bigr\}
\]
(cf.\ \cite{Dasgupta:1999ss,Gukov:1999ya}).  

The flux contributes to the M2-brane tadpole via
\[
  \frac{1}{2} \int_{X} G \wedge G = \, Q(G,G),
\]
and tadpole cancellation imposes the constraint\footnote{We have rescaled everything by a factor of 2 as it is more natural to impose on cohomology}
\begin{equation}
   \frac{\chi(X)}{24} 
      - \frac{1}{2} \int_{X} G \wedge G = N_{\mathrm{M2}} \in \mathbb{Z}_{\geq 0}.
   \label{eq:tadpole-bound}
\end{equation}
In particular, for primitive self-dual fluxes the right-hand side of \eqref{eq:tadpole-bound} is non-negative and the quantity
\[
   L(G) :=  Q(G,G)
\]
can be interpreted as the length (or norm) of the flux vector. Vacua are only allowed for classes with
\[
   L(G) \leq \frac{\chi(X)}{24},
\]
and we shall refer to Hodge classes with small $L(G)$ as short Hodge cycles.

\medskip

A second important dichotomy in our setting is whether the flux is strictly integral or half-integral. This is controlled by the parity of $c_2(X)$ in \eqref{eq:flux-quantization}: if $c_2(X)$ is even, then $G$ is integral, while for odd $c_2(X)$ the physically allowed classes live in a coset of the integral lattice, cf.\ \cite{Bena_2021,Dasgupta:1999ss}.

\subsection{Diagonal/symmetric points in moduli space}

For hypersurfaces it is convenient to distinguish special loci in complex structure moduli space where the defining equation enjoys enhanced symmetry. Concretely, suppose that $X$ is realized as a hypersurface in projective space with defining polynomial
\[
   P(x) = \sum_\alpha a_\alpha\, m_\alpha(x),
\]
where the $m_\alpha$ are monomials and $a_\alpha$ are complex parameters. Let $\Gamma$ be a finite group of automorphisms acting on the ambient space and preserving $X$.

\begin{definition}
A point $T\in B$ is called diagonal (or symmetric) if the defining polynomial $P$ can be chosen so that it is invariant under a fixed finite group $\Gamma$ of automorphisms. We refer to the corresponding locus in $B$ as the symmetric locus for the group~$\Gamma$.
\end{definition}

Physically, such symmetric loci are natural candidates for flux vacua: they are characterized by enhanced discrete automorphism groups, and it is often possible to choose fluxes that are themselves invariant under $\Gamma$. This is one of the organizing principles behind ``symmetric fluxes with small tadpoles'' in \cite{Coudarchet:2023mmm}, where symmetric fluxes are shown to give large families of vacua with a fixed (and relatively small) tadpole contribution, essentially independent of the number of moduli.

\subsection{The tadpole conjecture and finiteness results}

Motivated by the observation that stabilizing many complex structure moduli tends to be ``expensive'' in tadpole charge, Bena et al.\ formulated in \cite{Bena_2021} the tadpole conjecture. In the present language, this can be stated (schematically) as follows.

\begin{conjecture}[Tadpole conjecture, heuristic form]
  Let $G \in H_{2,2;\mathrm{prim}}(X)\cap H_4(X,\mathbb{Z})$ be a primitive integral Hodge class, and suppose that $aG$ is a general Hodge cycle
  \[
     a G = \sum_{i=0}^s n_i [Z_i],
  \]
  for some $s \in \mathbb{N}$, algebraic cycles $Z_i$ and non-zero integer $a$. Then for every point $T$ in moduli space the tadpole contribution satisfies
  \[
     Q_T(aG,aG) \;\gtrsim\; \frac{\chi(X_T)}{24}.
  \]
  Moreover, there exists a constant $C>0$, a priori depending on the family, such that the codimension of the Hodge locus associated to $aG$ obeys
  \[
     \mathrm{codim}_B\bigl(V_{aG}\bigr) 
     \;\leq\; C \cdot  Q(aG,aG).
  \]
\end{conjecture}

The refined tadpole conjecture further proposes that the constant $C$ is universal, with $C=4$, see for instance \cite{becker2024tadpoleconjecturenongeometricbackgrounds} for recent discussions.  

From a purely Hodge-theoretic perspective, there are now strong finiteness results that complement these conjectural bounds. In particular, Bakker, Grimm, Schnell and Tsimerman show in \cite{BakkerGrimmSchnellTsimerman:2023} a far-reaching finiteness theorem for self-dual classes in polarized integral variations of Hodge structure. Applied to the weight-four variation
\[
   H^4 \longrightarrow B
\]
arising from the family $X_T$ where $T$ parametrizes the complex structure moduli, their main result implies that for each fixed integer $q\geq 1$ the set of pairs
\[
   (T, v), \qquad T \in B, \; v \in H^4_T
\]
such that $v$ is integral, self-dual ($C_T v = v$) and $Q_T(v,v) = 2q$ is finite over $B$. In particular, for each fixed tadpole contribution
\[
   L = \frac{1}{2} Q_T(G,G)
\]
there are only finitely many self-dual integral Hodge classes with $L(G)=L$ in the entire family. This provides a rigorous Hodge-theoretic finiteness statement for ``self-dual short fluxes'' and underlies recent work on counting flux vacua from a mathematical viewpoint.

\subsection{Symmetric loci and short Hodge cycles}

The previous discussion suggests that short general Hodge cycles are rather constrained. At the same time, there is growing evidence that such cycles tend to live on special, highly symmetric loci in moduli space.

On the IIB side, Coudarchet et al.\ \cite{Coudarchet:2023mmm} showed that imposing invariance of the flux under a discrete symmetry of the prepotential allows one to construct large families of flux vacua with small tadpole, even in models with many complex structure moduli. This points to a general pattern: symmetry drastically reduces the effective dimension of the flux lattice while keeping enough degrees of freedom to stabilize many moduli, and the resulting symmetric fluxes tend to have relatively small length $L(G)$.

On the M-theory side, after investigating in detail the Fermat sextic fourfold in \cite{braun2024gfluxgeneralhodgecycles} using Griffiths residues, one can construct bases of integral Hodge cycles explicitly. In section~5.7, the action of finite symmetry groups 
\[
   \Gamma \subset \mathrm{Aut}(X)
\]
is used to descend symmetric fluxes to quotients $X/\Gamma$. For fluxes invariant under $\Gamma$, the self-intersection number on the quotient is effectively divided by $|\Gamma|$, while the Euler characteristic of a crepant resolution of $X/\Gamma$ is typically of the same order as $\chi(X)$ (see eqs.\ (5.57)–(5.59) and the surrounding discussion in \cite{braun2024gfluxgeneralhodgecycles}). As a consequence, the same pattern of complex-structure stabilization can often be realized on the quotient with a significantly reduced tadpole cost.

Although the quotients $X/\Gamma$ are in general singular and the corresponding fluxes may only be properly defined on the singular model (with fractional contributions interpreted in terms of M2-branes at orbifold singularities), these results provide strong evidence that symmetric loci in moduli space are natural habitats for short general Hodge cycles.
. 
In the following we use maximally symmetric loci as a controlled computational sector.  This choice is motivated by previous examples in which symmetry-invariant fluxes give relatively small tadpole charge \cite{Coudarchet:2023mmm}, and by the fact that the period matrices and Hodge-locus equations simplify substantially at Fermat points.  We do not assume, and do not prove, that the globally shortest physically admissible Hodge cycles must occur in this sector.  Instead, our computations should be read as an explicit test of the expectation that symmetry can lower the tadpole cost of moduli stabilization.

One of the lessons of the weighted examples below is that this expectation must be refined.  A flux may be symmetric under automorphisms of the hypersurface while its associated period data have a non-uniform Galois orbit.  In such cases, the integral reconstruction of the corresponding self-dual class can involve additional middle-cohomology components, and the resulting norm can be significantly larger than suggested by the naive symmetry argument.

\section{Notations and algorithm}

In this section we fix the notation for the class of hypersurfaces we will work with and recall the description of their periods following Movasati's books \cite{hodgetheory} I and II. Throughout we insist on the following condition.

\medskip

\noindent\textbf{Fermat / tame condition.}
Let $v_0,\dots,v_n$ be positive integers with $\gcd(v_0,\dots,v_n)=1$, and let
\[
   \P_{v_i} = \P(v_0,\dots,v_n)
\]
be the weighted projective space with these weights. We fix a positive integer $d$ such that
\[
   v_i \mid d \quad\text{for all } i,
\]
and we set
\[
   d_i := \frac{d}{v_i} \in \Z_{>0}.
\]
We say that a weighted homogeneous polynomial $g$ of degree $d$ is of Fermat type if, up to scaling of the coordinates, it is of the diagonal form
\[
   g(x) = x_0^{d_0} + x_1^{d_1} + \cdots + x_n^{d_n},
\]
and has an isolated singularity at the origin in $\C^{n+1}$.  

\medskip

In the applications we have $n=5$, so that the hypersurface will be a Calabi--Yau fourfold, but for the moment we keep $n$ general.

Let $D \subset \P_{v_i}$ be the projective hypersurface defined by a tame polynomial $g$ as above. Following Movasati (see \cite[§15]{hodgetheory}), we consider the polynomial
\[
    \tilde{g}(x_0,\dots,x_n) := g(x_0^{v_0},x_1^{v_1},\dots,x_n^{v_n})
\]
and the corresponding hypersurface
\[
    \tilde{D} := \{\,\tilde{g}=0\,\} \subset \P^n.
\]
The finite abelian group
\[
   G_{v_i} := \mu_{v_0} \times \cdots \times \mu_{v_n}
\]
of $v_i$-th roots of unity acts diagonally on $\tilde{D}$, and the quotient is $D$.

\begin{definition}
   The de Rham cohomology of $D$ is defined as the $G_{v_i}$-invariant part of the de Rham cohomology of $\tilde{D}$:
   \[
      H^k_{dR}(D) := H^k_{dR}(\tilde{D})^{G_{v_i}}.
   \]
\end{definition}

\begin{remark}
	The previous definition follows from the \textbf{Cartan model} for the equivariant cohomology for the $D=\tilde D/{G_{v_i}}$. Namely, the Cartan complex is 
	$$\Omega^\ast_{G_{v_i}}(\tilde D)=(S(\mathfrak{g^*})\otimes\Omega^\ast(\tilde D))^{G_{v_i}}=(\Omega^ \ast(\tilde D))^{G_{v_i}},$$
    where $\mathfrak{g}=\text{Lie}(G)=0$ and $S(\mathfrak{g}^*)$ is the symmetric algebra of the dual Lie algebra $\mathfrak{g}^*$,
	and the Cartan differential evaluated in $\omega:\mathfrak{g}\to \Omega^\ast(\tilde D)$  is $$(d_C\omega)(v)=d_{dR}(\omega(v))-\imath_{v_X}(\omega(v))=d_{dR}\omega, \text{ where } v\in \mathfrak{g}.$$
 Thus,  $H_{G_{v_i}}^k(\tilde D)$ is the $G_{v_i}$-equivariant part of $H_{dR}(\tilde D)$.
\end{remark}

We will be interested in the primitive middle cohomology $H^4_{dR}(D)_{\prim}$ ($n=5$ in our case) and its Hodge decomposition. Movasati constructs an explicit basis of $H^4_{dR}(\tilde{D})$ in terms of residues of meromorphic differential forms attached to monomials; we briefly recall the part of the construction that we need.

Let
\[
   \beta = (\beta_0,\dots,\beta_n), \qquad 0 \le \beta_i \le d_i-2,
\]
be a multi-index. Denote by $x^\beta = x_0^{\beta_0}\cdots x_n^{\beta_n}$ the corresponding monomial and by $\Omega$ the standard volume form on $\P^n$. As in \cite[§15--§16]{hodgetheory} one considers the meromorphic forms
\[
   \frac{x^\beta \,\Omega}{g^k}
\]
with appropriate $k$, and their residues along $D$:
\[
   \omega_\beta := \operatorname{Res}_{D}\left(\frac{x^\beta \,\Omega}{g^k}\right) \in H^{n-1}_{dR}(D).
\]
After imposing a suitable combinatorial condition on $\beta$, the classes $\omega_\beta$ form a basis of the primitive part of $H^{n-1}_{dR}(D)$ compatible with the Hodge filtration (see \cite{hodgetheory} for details). For the rest of this section we fix such a set of indices and write simply
\[
   \{\omega_\beta\}_\beta
\]
for this Hodge-compatible basis of $H^{n-1}_{dR}(D)_{\prim}$.

\subsection{Working in (Co)homology}

\subsubsection{Periods and the cyclotomic field}

Let $\delta$ be a Hodge cycle on $D$. The main structural input from Movasati that we use is the cyclotomic nature of the periods.

\medskip

\noindent\textbf{Theorem (Movasati, \cite[Thm.~16.1]{hodgetheory}).}
Let $d$ be the degree of $g$ as above, and $\zeta_d$ a primitive $d$-th root of unity. Then for every Hodge cycle $\delta$ and every basis element $\omega_\beta$ we have
\[
   \frac{1}{(2\pi i)^{n}}\int_\delta \omega_\beta \in \Q[\zeta_{d}].
\]
In particular, all normalized periods are in the cyclotomic field $K:=\Q[\zeta_{d}]$.

\medskip

In the diagonal case (Fermat) in which we are interested, Proposition~15.1 of \cite{hodgetheory} gives an explicit formula for the periods in terms of $\Gamma$-functions and roots of unity. Adapting it to the weighted situation (with the tame condition $v_i\mid d$) one obtains, for vanishing cycles $\delta_{\beta'}$ indexed in a compatible way,
\begin{align}\label{periods_zeta}
\int_{\delta_{\beta'}}\omega_{\beta}
 &= \frac{1}{\prod_{i=0}^{n} d_i \cdot \frac{n}{2}! \,(2\pi i)}\,
    \prod_{i=0}^{n}
      \Bigl(
        \zeta_{d_i}^{(\beta_i+1)\beta'_i}
        \bigl(\zeta_{d_i}^{\beta_i+1}-1\bigr)\,
        \Gamma\!\Bigl(\frac{\beta_i+1}{d_i}\Bigr)
      \Bigr) \\
 &=: z_u \, z_\beta \, Z(\beta,\beta').
\end{align}
We separate the period into a transcendental normalization and a cyclotomic factor,
\[
   \int_{\delta_{\beta'}} \omega_\beta
   =
   z_{\rm univ}\, z_\beta\, Z(\beta,\beta') ,
\]
where \(z_{\rm univ}\) depends only on the degree and the weights, \(z_\beta\) is the product of Gamma factors depending on the residue form, and
\[
   Z(\beta,\beta')\in \mathbb Q(\zeta_d)
\]
is the root-of-unity factor.  In the lattice computations below, the arithmetic information used to impose the rationality and integrality constraints is carried by the cyclotomic matrix \(Z(\beta,\beta')\).  The Gamma factors give the Hodge normalization of the chosen residue basis and are kept separate from the cyclotomic reconstruction.

Furthermore, the intersection between residue forms is modulo the Jacobian ideal. In the context of Fermat varieties, this means that two residue forms intersect only when they are complex conjugate of each other : the exponents of the numerator of $\omega_{\beta_1}$ and $\omega_{\beta_2}$ must correspond to the Hessian ideal. In this same context, any residue form which do not meet this requirement will lie in the Jacobian ideal.

Lastly, knowing this, consider a general $G_4$ flux : $G_4=\sum_i a_i \omega_i$. Then the tadpole contribution can be written symbolically as $\frac{1}{2}G_4\wedge G_4:=\frac{1}{2}G_4\cdot G_4=\frac{1}{2}\sum_i 2\cdot (a_i\omega_i,\overline{a_i\omega_i})$, with the factor of 2 coming from the symmetry and the non-complex conjugate residue forms being 0 as above. Hence once we know the intersection form of the $\omega_i$ we can conclude by computing the shortest vector since the $\frac{1}{2}$ factor cancels out.

To provide an explicit example of the last two points, consider the 4 following forms for the degree 36 Fermat variety in $\P_{1,1,1,9,12,12}$ that we will use later:
\begin{align*}
 \omega_0=[24, 24, 24, 0, 0, 0], \\
 \omega_1=[16, 16, 16, 0, 1, 1], \\
 \omega_4=[18, 18, 18, 2, 0, 0], \\
 \omega_5=[10, 10, 10, 2, 1, 1]
\end{align*}

We take for example $\omega_0$ and $\omega_5$. Those two residue forms are complex conjugate of each other because $[24, 24, 24, 0, 0, 0]+[10, 10, 10, 2, 1, 1]=[34,34,34,2,1,1]$ which corresponds to the Hessian ideal in this case. However $\omega_0$ and $\omega_1$ do not : $[24, 24, 24, 0, 0, 0]+[16, 16, 16, 0, 1, 1]=[40,40,40,0,1,1]$. For this example, notice that we have an entry of 40, and hence it lies in the Jacobian ideal because $(x_0^{35})\subset(x_0^{40})$ rendering this intersection 0.

Now let us construct an example of a $G_4$ flux using those 4 residue forms to give an example of the last point:
\begin{align*}
    G_4=\omega_0+\omega_1+\omega_4+\omega_5
\end{align*}

We compute its length, which by definition is:
\begin{align*}
    \frac{1}{2}\int_X G_4\wedge G_4:=\frac{1}{2}G_4\cdot G_4=\frac{1}{2}\int_X(\omega_0+\omega_1+\omega_4+\omega_5)\wedge(\omega_0+\omega_1+\omega_4+\omega_5)
\end{align*}
We expand the wedge product, notice that the only pairs of complex conjugate are $(\omega_0,\omega_5)$ and $(\omega_1,\omega_4)$. Those pairs appear 4 times, and the rest is 0 because the lie in the Jacobian ideal.

We end up with:
\begin{align*}
    \frac{1}{2}G_4\cdot G_4=\frac{1}{2}\int_X(\omega_0\wedge \omega_5 + \omega_5 \wedge \omega_0 + \omega_1 \wedge \omega_4 + \omega_4 \wedge \omega_1)
\end{align*}
But those forms are in $H^{2,2}$. Thus we have:
\begin{align*}
    \omega_0\wedge \omega_5 = (-1)^4 \omega_5 \wedge \omega_0 = \omega_5 \wedge \omega_0
\end{align*}

Thus:
\begin{align}
    \frac{1}{2}G_4\cdot G_4 = \frac{1}{2} \cdot 2 \cdot \int_X( \omega_0 \wedge \omega_5 + \omega_1\wedge \omega_4)
\end{align}
The factors of 2 and $1/2$ cancel each other and we are left simply with the intersection form.

\subsubsection{Galois action in cohomology}

Since we will work in cyclotomic fields $\Q[\zeta_{d}]$ over generalized Fermat varieties of degree $d$, we have to take into account the Galois action that comes from the Galois group.

Indeed from \cite{Duque_Franco_2023}, we know how the Galois group acts on residue forms. Let \(\alpha=(\beta_0+1,\ldots,\beta_5+1)\) denote the character associated with the residue form \(\omega_\beta\).  For each Galois embedding
\[
   \sigma_a:\zeta_d\mapsto \zeta_d^a,\qquad (a,d)=1,
\]
the cyclotomic part of the period matrix transforms by sending the character \(\alpha\) to \(a\alpha\) modulo the appropriate degrees.  The Hodge type of the corresponding eigenspace is then determined by the weighted degree of this transformed character.

In the ordinary sextic case the relevant Galois orbit of a \((2,2)\)-character remains inside the \((2,2)\)-sector together with its complex conjugate.  We call such an orbit uniform.  In weighted Fermat examples this need not happen: a \((2,2)\)-character can have Galois conjugates of type \((3,1)\), \((1,3)\), or \((4,0)+(0,4)\).  We refer to this situation as a non-uniform Galois orbit.

\subsubsection{From periods to the integral lattice}\label{Algo}

In general there is no canonical way to choose a basis of vanishing cycles such that the period matrix is already adapted to a $\Z$-basis of $H^n(D,\Z)_{\prim}$. In particular, for the weighted Fermat examples we consider (as opposed to the ordinary sextic) there is no simple normalization that makes the lattice structure manifest. Instead we proceed in several steps.

In what follows we use the convention
\[
P_{\beta i}=\int_{\delta_\beta}\omega_i,
\]
so that rows are indexed by vanishing cycles and columns by residue forms.
With this convention \(P\) is an \(N\times r\) matrix and the equation
\[
\Psi X=P
\]
is dimensionally consistent. This differs from the convention
\(\Pi_{i\beta}\) by transposition. The algorithm is the following:

\begin{enumerate}
\item Choose a symmetry-invariant set of residue forms \(\{\omega_i\}\) spanning the sector of \(H^{2,2}\) under consideration.

\item Compute the cyclotomic period matrix
\[
   \Pi_{\beta'i}=\int_{\delta_{\beta'}}\omega_i
\]
on the chosen basis of vanishing cycles.

\item Expand the cyclotomic entries in a fixed \(\mathbb Q\)-basis of \(\mathbb Q(\zeta_d)\).  This replaces \(\Pi\) by a rational matrix \(P\).

\item Let \(\Psi\) be the intersection matrix of the vanishing cycles.  Solving
\[
   \Psi X=P
\]
expresses the dual cohomology classes in the vanishing-cycle basis.

\item Let \(V_{\mathbb Q}\subset H_4(D,\mathbb Q)\) be the rational row span generated by \(X^T\).  The integral lattice used in the enumeration is the saturation
\[
   \Lambda = V_{\mathbb Q}\cap H_4(D,\mathbb Z).
\]
It is computed by using SAGEMATH routine.

\item The Gram matrix on \(\Lambda\) is
\[
   Q_\Lambda = B\,\Psi\,B^T ,
\]
where the rows of \(B\) form an integral basis of \(\Lambda\).
\end{enumerate}

When the Galois orbit is non-uniform, \(Q_\Lambda\) need not be positive definite, because the rational reconstruction may include contributions from \(H^{3,1}\oplus H^{1,3}\) and \(H^{4,0}\oplus H^{0,4}\).  We therefore decompose the resulting lattice according to the Hodge signature and retain the positive self-dual sector relevant for flux enumeration.
The insistence on the Fermat/tame condition (each weight dividing the degree) ensures that the period formula \eqref{periods_zeta} has the simple multiplicative structure above and that the cyclotomic field $K'=\Q[\zeta_{d}]$ captures all the arithmetic of the periods. Without this tameness the explicit control on $Z(\beta,\beta')$ breaks down, and the algorithm would no longer be directly applicable.

Notice that the lattice saturation is necessary. Let \(H_{\mathbb Z}\) be the torsion-free integral middle homology lattice, with
intersection matrix \(\Psi\), and let
\[
V_{\mathbb Q} \subset H_{\mathbb Q}
\]
be the rational subspace cut out by the cyclotomic period constraints corresponding to
the chosen symmetric \(H^{2,2}\)-forms. The integral lattice used for flux enumeration is
\[
\Lambda := V_{\mathbb Q} \cap H_{\mathbb Z}.
\]

If \(B_{\mathbb Q}\) is any rational row basis for \(V_{\mathbb Q}\), then clearing
denominators followed by Hermite or Smith normal form computes the saturated lattice
\(\Lambda\), independently of the initial rational basis.

Once the lattice is formed, further steps are needed. Crucially, we rely on the fact that the modules $H^{4,0}, H^{3,1}, H^{2,2}, H^{1,3}, H^{0,4}$ are in a direct sum, and that the intersection form on $H^{3,1}$ and $H^{1,3}$ is negative definite. This allows us to identify which components of the Gram matrix correspond to which block in this decomposition by then choosing an appropriate basis.

We provide an implementation for our example of degree 36 in \url{https://github.com/hugofortin/TCIWPS}.

\subsubsection{Further remarks and comparison with Movasati's algorithm}

The first point to notice is to carefully pick the affine coordinate chart in a weighted projective space: there should not be residual $\Z^n$ actions. Fortunately for us, all varieties studied here have a coordinate with weight 1, so they are perfectly good to define the affine plane we work in.

Regarding the computational cost, we have to keep in mind the dimensions of the full problem and the state of the art when it comes to short-vector enumeration, nevermind the issue of checking if the vectors are indeed general Hodge cycles. The state of the art is in the low hundreds in terms of dimensions, which for Fermat-type polynomial is far below the actual dimension of $H^{2,2}$. Thus working with the full lattice is unrealistic.

A direct kernel computation on the full vanishing-cycle lattice is not the
appropriate computational scale for the present problem. The target is not the
full \(H^{2,2}\)-lattice but the maximally symmetric sector. The kernel construction aims to build the full lattice and is more efficient if that were the case, however we are only interested in symmetric sub-lattices and the kernel construction can only produce the full lattice.

Let us also explain why the intersection form obtained from the two constructions
agrees on the actual \(H^{2,2}\)-sector. Let
\[
H_{\mathbb Z}:=H_4(D,\mathbb Z)_{\mathrm{prim}}/\mathrm{tors},
\qquad
H_{\mathbb Q}:=H_{\mathbb Z}\otimes_{\mathbb Z}\mathbb Q,
\]
and let \(\Psi\) be the intersection matrix in the chosen vanishing-cycle basis. The
intersection pairing is the fixed bilinear form
\[
Q(u,v)=u\,\Psi\,v^T,
\qquad u,v\in H_{\mathbb Q}.
\]
Thus neither construction produces a new pairing: both only produce subspaces of the
same ambient homology space \(H_{\mathbb Q}\), and the intersection numbers are obtained
by restricting \(Q\).

Movasati's construction gives the Hodge subspace directly as
\[
S_{\mathbb Q}
=
H_{\mathbb Q}\cap H^{2,2}
=
\ker(\widetilde P_2)
\]
inside the chosen symmetric sector, where \(\widetilde P_2\) is the rational expansion of
the period matrix for \(H^{4,0}\oplus H^{3,1}\). If the rows of \(Y\) form a basis of this
kernel, then the corresponding Gram matrix is
\[
G_Y=Y\Psi Y^T.
\]

On the other hand, the construction
\[
\Psi X=\widetilde P_1
\]
starts from the period matrix of the selected \((2,2)\)-forms. Since our convention is that
rows are cycles and columns are forms, the rational expansion is performed column-wise and
the columns of \(X\) are the reconstructed rational homology classes. Equivalently, one may
work with the row span of \(X^T\). This gives a rational Galois-completed space
\[
V_{\mathbb Q}^{\mathrm{raw}}
=
\operatorname{ColSpan}_{\mathbb Q}(X)
\subset H_{\mathbb Q}.
\]
In the non-uniform Galois case this space may contain non-\((2,2)\) directions, so one must
retain only the Hodge part
\[
V_{\mathbb Q}^{\mathrm{Hdg}}
=
V_{\mathbb Q}^{\mathrm{raw}}\cap \ker(\widetilde P_2).
\]

If the selected \((2,2)\)-forms generate the full symmetric \(H^{2,2}\)-sector, then
\[
V_{\mathbb Q}^{\mathrm{Hdg}}
=
S_{\mathbb Q}.
\]
Hence the two constructions give the same rational subspace of \(H_{\mathbb Q}\). Therefore
their Gram matrices differ only by a rational change of basis. Indeed, if \(X_{22}\) and
\(Y\) are two rational bases of \(S_{\mathbb Q}\), then for some
\(T\in \operatorname{GL}_r(\mathbb Q)\),
\[
X_{22}=TY,
\]
and so
\[
X_{22}\Psi X_{22}^T
=
T\,Y\Psi Y^T\,T^T.
\]
After saturation, both constructions give the same integral lattice
\[
\Lambda
=
S_{\mathbb Q}\cap H_{\mathbb Z}.
\]
Thus any two integral bases \(B_X\) and \(B_Y\) of \(\Lambda\) differ by some
\(U\in \operatorname{GL}_r(\mathbb Z)\), and their Gram matrices satisfy
\[
B_X\Psi B_X^T
=
U\,B_Y\Psi B_Y^T\,U^T.
\]
Therefore the two constructions give isometric integral quadratic lattices on the true
\(H^{2,2}\)-sector. In particular, the intersection numbers, norms, shortest vectors, and
tadpole contributions agree up to an integral change of basis.

In the examples computed, after the cyclotomic normalization the matrices have
integral entries, and the subsequent saturation computation returns index one.
The latter is an additional exact check: integrality of the generators alone
does not imply saturation. In general for general varieties it is necessary to pass from the rational subspace to the associated integral lattice. We use saturation in the standard $\mathbb{Z}$-module sense: if
$L \subset \mathbb{Z}^n$ is a lattice, then
\[
L^{\mathrm{sat}}
=
\left(L \otimes_{\mathbb{Z}} \mathbb{Q}\right)\cap \mathbb{Z}^n
=
\left\{
x\in \mathbb{Z}^n \;\middle|\;
\exists\, m\in \mathbb{Z}_{>0}\ \text{such that}\ mx\in L
\right\}.
\]
Computationally, this saturation can be obtained using Hermite or Smith normal form
methods for finitely generated $\mathbb{Z}$-modules~\cite{Cohen1993ComputationalANT}.

There are two caveats to the algorithm: the first one being that we have to find a suitable integral unimodular change of basis to make explicit the different blocks that are in direct sum. For the small Gram matrices appearing in the examples, we determine such an integral change of basis by an exact search over unimodular transformations subject to the required block constraints. However, for the general case of the full $H^{2,2}$ lattice, Movasati's constraint-first algorithm is more suitable.

Lastly, the second caveat is that we do not explicitly compute the primitive vertical part, as residue forms can only span the primitive horizontal part. In our examples, this is only a problem for the degree 36 example where the dimension of the primitive vertical part is 1, in all the rest it is 0. In principle, the intersection number of the vertical part is computable using the results of \cite{Villaflor_Loyola_2024} and mirror symmetry. Furthermore, using the orthogonal decomposition of the middle cohomology into horizontal,
vertical and remaining pieces, the primitive vertical lattice is orthogonal to
the horizontal residue lattice. Hence for
\[
G=G_H+G_V
\]
one has
\[
\frac12\int_X G\wedge G
=
\frac12\int_X G_H\wedge G_H
+
\frac12\int_X G_V\wedge G_V.
\]
Since the primitive vertical \((2,2)\)-sector is self-dual and positive for the
polarization used here, adding \(G_V\) cannot decrease the tadpole contribution.
Moreover, \(G_V\) is algebraic and does not enter the horizontal variation of
\(\Omega\), hence it does not change the complex-structure Hessian.
The results we found is that the horizontal part already overshoots the tadpole bound, so that adding the vertical part would only make things worse with respect to the bound. For the sake of model building this vertical part is however crucial, as highlighted in \cite{Braun_2015}, so there is possibly some further tension here.

\subsection{Galois completion and the primitive condition}

A subtle point in the weighted examples is the passage from cyclotomic period data to an integral lattice over \(\mathbb Q\).  Expanding the period matrix in a \(\mathbb Q\)-basis of \(\mathbb Q(\zeta_d)\) amounts to taking the Galois completion of the corresponding eigenspaces.  If the Galois orbit of a \((2,2)\)-character is non-uniform, this completion can contain components of Hodge type different from \((2,2)\).  Consequently the intermediate Gram matrix obtained from vanishing cycles may have indefinite signature.

However we know that the Hodge decomposition is an orthogonal sum. Thus right away we can identify the $H^{3,1}+H^{1,3}$ block without effort since it should be the only block with negative definite signature. However, the identification for the remaining positive blocks is more subtle. Fortunately, in the examples we study, we can identify the $H^{4,0}+H^{0,4}$ block explicitly when needed, and thus conclude about the minimal length of a primitive general Hodge cycle.

The vertical primitive part of the middle cohomology does not affect the Hodge-locus condition used to test whether a symmetric flux is general, since these classes are algebraic. Moreover, its intersection form is positive on the self-dual sector, so adding vertical primitive components cannot lower the minimal norm found in the horizontal symmetric sector. Therefore the lower bound on the norm of primitive general Hodge cycles used below is unaffected by the vertical sector.

\section{Examples of hypersurfaces in weighted projective spaces}

\paragraph{Scope of the computation.}
All results below concern the maximally symmetric flux sector. Within this sector, the degree-36, degree-12, and degree-8 examples verify the tadpole bound for primitive general Hodge cycles. We do not claim a global classification of all fluxes in the full middle cohomology.  

In this section we illustrate the general discussion with concrete examples of tame generalized Fermat hypersurfaces in weighted projective spaces. Conceptually, these should be compared with the Fermat sextic fourfold studied in \cite{braun2024gfluxgeneralhodgecycles}: there the ambient space is the ordinary projective space $\P^5$, the hypersurface is of degree~6, and one has $h^{1,1}=1$. As a consequence the primitive middle cohomology is essentially “purely horizontal’’ and the flux lattice is relatively simple.

In the weighted situation considered here, several new features appear:

\begin{itemize}
  \item the ambient is a weighted projective space $\P_{v_i}$, with at least one weight $v_i>1$;
  \item $h^{1,1}$ is generally larger than~$1$, so one has a non-trivial vertical part of $H^{2,2}$ in addition to the horizontal one;
  \item the Euler characteristic $\chi(X)$ is typically much larger, which pushes the tadpole bound $\chi(X)/24$ upward;
  \item the tame (Fermat) condition
  \[
      v_i \mid d \quad\text{for all }i
  \]
  ensures that the method of Section~2 applies: Movasati’s period formulas, the cyclotomic description of periods, and the algorithm described in \ref{Algo} for constructing an integral lattice all go through essentially unchanged.
  \item the Galois action is typically non-trivial.
\end{itemize}

We will see that, from the point of view of the tadpole conjecture, all examples we study verify the conjecture. Furthermore, we observe that typically a uniform Galois action leads to short cycles.

Importantly, we use the same enumeration algorithm as for the sextic case, that is SAGEMATH implementation of Fincke-Pohst, and hence all of our results are exact and the numbers computed indeed corresponds to the shortest vectors.

\subsection{A degree-36 weighted Fermat example with \(h^{1,1}=11\)}

Our first example is the following degree~36 generalized Fermat hypersurface in
\[
   \P_{1,1,1,9,12,12}
\]
which is elliptically fibered:
\begin{align}
    X_0 := x_0^{36}+x_1^{36}+x_2^{36}+x_3^4+x_4^3+x_5^3.
\end{align}
The polynomial is of Fermat/tame type: each weight divides the degree,
\[
   1\mid 36,\quad 9\mid 36,\quad 12\mid 36,
\]
and $g$ has an isolated singularity at the origin. Hence the setup of Section~2 applies verbatim.

We denote its group of automorphisms by
\[
   G_0 = \bigl(\Z_{36}^3\times\Z_4\times\Z_3^2\bigr) \rtimes (S_3\times S_2)
\]
modulo the diagonal subgroup. The discrete symmetry is considerably larger than for the sextic in $\P^5$, and in particular the maximally symmetric point in complex-structure moduli space is no longer diagonal in the usual sense: there is a non-trivial mixing among the coordinates induced by the permutation part of $G_0$.

The relevant Hodge numbers of $X_0$ are
\begin{align}
   h^{1,1}(X_0) = 11,\qquad
   h^{2,2}(X_0) = 9564,\qquad
   h^{1,3}(X_0) = 2369,
\end{align}
with tadpole bound
\[
   \frac{\chi(X_0)}{24} = 597.
\]
Thus, compared to the sextic fourfold, both the number of moduli and the middle cohomology are much larger, with a huge space of $H^{2,2}$ classes. The second Chern class $c_2(X_0)$ is even, so the flux quantization condition
\[
   G_4+\frac{1}{2}c_2(X_0)\in H^4(X_0,\Z)
\]
allows integral fluxes $G_4$.

We are interested in the fully symmetric sector of the moduli space, which leaves us with:
\[
   h^{1,3} = 5,\qquad h^{2,2} = 6,
\]
so the symmetric Hessian matrix in moduli space is $5$-dimensional, and the symmetric fluxes live in a $6$-dimensional subspace of $H^{2,2}$.

The symmetric Hodge classes can be described as follows. We consider monomials
\[
   x_0^{\beta_0}\cdots x_5^{\beta_5}
\]
of total weighted degree $36$, and impose invariance under the diagonal part of $G_0$ modulo the Jacobian ideal. This yields a finite set of exponent vectors
\[
   \beta = (\beta_0,\dots,\beta_5)
\]
which we interpret, as in Section~2, as indexing a basis $\{\omega_i\}$ of the $G_0$–invariant part of $H^{2,2}(X_0)$. 

Invariant $H^{2,2}$ forms $\omega_i$:
\begin{align*}
 \omega_0=[24, 24, 24, 0, 0, 0], \\
 \omega_1=[16, 16, 16, 0, 1, 1], \\
 \omega_2=[21, 21, 21, 1, 0, 0],\\
 \omega_3=[13, 13, 13, 1, 1, 1], \\
 \omega_4=[18, 18, 18, 2, 0, 0], \\
 \omega_5=[10, 10, 10, 2, 1, 1]
\end{align*}

The Galois action is given, for example for the $\omega_0$ form by : 
\begin{align*}
1    | [25, 25, 25, 1, 1, 1]  \\
5    | [17, 17, 17, 1, 2, 2]  \\
7    | [31, 31, 31, 3, 1, 1]  \\
11   | [23, 23, 23, 3, 2, 2]  \\
13   | [1, 1, 1, 1, 1, 1]     \\
17   | [29, 29, 29, 1, 2, 2]  \\
19   | [7, 7, 7, 3, 1, 1]     \\
23   | [35, 35, 35, 3, 2, 2]  \\
25   | [13, 13, 13, 1, 1, 1]  \\
29   | [5, 5, 5, 1, 2, 2]     \\
31   | [19, 19, 19, 3, 1, 1]  \\
35   | [11, 11, 11, 3, 2, 2]  
\end{align*}
where the first number corresponds to the exponent of the Galois embedding, and the tuple corresponds to the character associated to the residue form, ie $\alpha=[\beta_0+1,\beta_1+1,\ldots]$. Notably,  the pair $\omega_0, \omega_5$ maps to forms in $H^{1,3}$ and $H^{4,0}$.

Invariant $H^{3,1}$ forms:
\begin{align*}
[12, 12, 12, 0, 0, 0], \\
[4, 4, 4, 0, 1, 1],\\
[9, 9, 9, 1, 0, 0], \\
[1, 1, 1, 1, 1, 1], \\
[6, 6, 6, 2, 0, 0]
\end{align*}

Leading to the Hessian matrix :
\[Hessian=
\begin{pmatrix}
\omega_5 & \omega_4 & \omega_3 & \omega_2 & \omega_1 \\
\omega_4 & 0        & \omega_2 & 0        & \omega_0 \\
\omega_3 & \omega_2 & \omega_1 & \omega_0 & 0        \\
\omega_2 & 0        & \omega_0 & 0        & 0        \\
\omega_1 & \omega_0 & 0        & 0        & 0
\end{pmatrix}
\]

The practical way to derive this Hessian is to consider $H^{3,1}$ and $H^{2,2}$ forms that sum up to the Hessian ideal. Take the first entry, for example :

$[12, 12, 12, 0, 0, 0]+[12, 12, 12, 0, 0, 0]+[10, 10, 10, 2, 1, 1]=[34,34,34,2,1,1]$.

Notice that for this particular Hessian matrix to have full rank, a condition is that $\omega_0$ is non-zero since the determinant is \[
\det(Hessian)
=
\omega_0^2
\left(
\omega_0^2\omega_5
-2\omega_0\omega_1\omega_4
-2\omega_0\omega_2\omega_3
+3\omega_1\omega_2^2
\right).
\]

From the point of view of the tadpole conjecture, however, this example behaves very differently from the sextic.

The Gram matrix we obtained is : 
{\tiny\begin{align}
   \left( 
\begin{array}{rrrrrrrrrrrrrrrrrr}
0 & 0 & 432 & -864 & -864 & 432 & 0 & 0 & 432 & -864 & -864 & 432 & 0 & -3888 & -3888 & 3888 & 1296 & 1296 \\
0 & 0 & 432 & 432 & 432 & -864 & 0 & 0 & 432 & 432 & 432 & -864 & 3888 & 0 & 0 & -3888 & -2592 & 1296 \\
432 & 432 & 0 & 0 & 432 & -864 & 432 & 432 & 0 & 0 & 432 & -864 & 2592 & -1296 & 0 & -3888 & -3888 & 3888 \\
-864 & 432 & 0 & 0 & 432 & 432 & -864 & 432 & 0 & 0 & 432 & 432 & -1296 & 2592 & 3888 & 0 & 0 & -3888 \\
-864 & 432 & 432 & 432 & 0 & 0 & -864 & 432 & 432 & 432 & 0 & 0 & 1296 & 1296 & 2592 & -1296 & 0 & -3888 \\
432 & -864 & -864 & 432 & 0 & 0 & 432 & -864 & -864 & 432 & 0 & 0 & -2592 & 1296 & -1296 & 2592 & 3888 & 0 \\
0 & 0 & 432 & -864 & -864 & 432 & 0 & 0 & 432 & -864 & -864 & 432 & 0 & -3888 & -3456 & 3024 & 432 & 1728 \\
0 & 0 & 432 & 432 & 432 & -864 & 0 & 0 & 432 & 432 & 432 & -864 & 3888 & 0 & 432 & -3456 & -2160 & 432 \\
432 & 432 & 0 & 0 & 432 & -864 & 432 & 432 & 0 & 0 & 432 & -864 & 3024 & -864 & 0 & -3888 & -3456 & 3024 \\
-864 & 432 & 0 & 0 & 432 & 432 & -864 & 432 & 0 & 0 & 432 & 432 & -2160 & 3024 & 3888 & 0 & 432 & -3456 \\
-864 & 432 & 432 & 432 & 0 & 0 & -864 & 432 & 432 & 432 & 0 & 0 & 432 & 1728 & 3024 & -864 & 0 & -3888 \\
432 & -864 & -864 & 432 & 0 & 0 & 432 & -864 & -864 & 432 & 0 & 0 & -2160 & 432 & -2160 & 3024 & 3888 & 0 \\
0 & 3888 & 2592 & -1296 & 1296 & -2592 & 0 & 3888 & 3024 & -2160 & 432 & -2160 & 15552 & -7776 & 3456 & -18576 & -22464 & 11232 \\
-3888 & 0 & -1296 & 2592 & 1296 & 1296 & -3888 & 0 & -864 & 3024 & 1728 & 432 & -7776 & 15552 & 15120 & 3456 & 11232 & -22464 \\
-3888 & 0 & 0 & 3888 & 2592 & -1296 & -3456 & 432 & 0 & 3888 & 3024 & -2160 & 3456 & 15120 & 15552 & -7776 & 3456 & -18576 \\
3888 & -3888 & -3888 & 0 & -1296 & 2592 & 3024 & -3456 & -3888 & 0 & -864 & 3024 & -18576 & 3456 & -7776 & 15552 & 15120 & 3456 \\
1296 & -2592 & -3888 & 0 & 0 & 3888 & 432 & -2160 & -3456 & 432 & 0 & 3888 & -22464 & 11232 & 3456 & 15120 & 15552 & -7776 \\
1296 & 1296 & 3888 & -3888 & -3888 & 0 & 1728 & 432 & 3024 & -3456 & -3888 & 0 & 11232 & -22464 & -18576 & 3456 & -7776 & 15552
\end{array}\right)
\end{align}}

It has signature (8,10) and so is indefinite.
After putting it in block diagonal form by brute forcing a search of an unimodular integral change of basis matrix, we find an 8 by 8 positive definite matrix. In particular this matrix has a 2 by 2 block and a 6 by 6 block. Using the standard result that $H^{2,2}$ and $H^{4,0}$+$H^{0,4}$ are in a direct sum we can identify the blocks accordingly. We obtain :
\begin{align}
G_{36}=
    \left(\begin{array}{rrrrrrrr}
864 & -432 & 0 & 0 & 0 & 0 & 0 & 0 \\
-432 & 864 & 0 & 0 & 0 & 0 & 0 & 0 \\
0 & 0 & 2592 & -432 & 0 & -864 & 0 & 1728 \\
0 & 0 & -432 & 864 & -432 & -432 & 432 & 432 \\
0 & 0 & 0 & -432 & 864 & -864 & 432 & 0 \\
0 & 0 & -864 & -432 & -864 & 3456 & -2160 & -2160 \\
0 & 0 & 0 & 432 & 432 & -2160 & 1728 & 1296 \\
0 & 0 & 1728 & 432 & 0 & -2160 & 1296 & 2592
\end{array}\right)
\end{align}
  
Now we have a problem of identifying which vector comes from which form. To remedy this problem, we can notice the following: the Galois action for the $\omega_0$ form contains the $H^{4,0}$ form and its complex conjugate, as well as $\omega_0$ and its complex conjugate $\omega_5$, and $\omega_1$ and its complex conjugate $\omega_4$. Furthermore, notice that the previous lattice we obtained for the full symmetric lattice has a deformed $432\cdot A_2$ block as well as a deformed $432\cdot E_6$ block which is indecomposable.

Recall that when we write a general $G_4$ flux and its intersection form, we write:
\begin{align*}
    \frac{1}{2}G_4\wedge G_4= \frac{1}{2}\sum_i 2\cdot a_i\omega_i\cdot \overline{a_i \omega_i}
\end{align*}

where the non-complex conjugate residue forms pairs automatically lie in the Jacobian ideal, which is modded out, and the factor of 2 comes from the symmetry.  Thus it is sufficient to study the exact norm rather than the half-norm associated with the 6 by 6 block of $G_{36}$. Note that

\begin{align}
    \left(\begin{array}{rrrrrr}
2592 & -432 & 0 & -864 & 0 & 1728 \\
-432 & 864 & -432 & -432 & 432 & 432 \\
-432 & 864 & -864 & 432 & 0 \\
-864 & -432 & -864 & 3456 & -2160 & -2160 \\
0 & 432 & 432 & -2160 & 1728 & 1296 \\
1728 & 432 & 0 & -2160 & 1296 & 2592
\end{array}\right)=432     \left(\begin{array}{rrrrrr}
6 & -1& 0 & -2 & 0 & 4 \\
-1 & 2 & -1 & -1 & 1 & 1 \\
0&-1& 2 & -2 & 1 & 0 \\
-2 & -1 & -2 & 8 & -5 & -5 \\
0 & 1 & 1 & -5 & 4 & 3 \\
4 & 1 & 0 & -5 & 3 & 6
\end{array}\right).
\end{align}
Moreover, the quadratic form associated with the matrix on the right side is 
\begin{align*}
q(x)=6x_1^2+2x_2^2+2x_3^2+8x_4^2+4x_5^2+6x_6^2-2x_1x_2-4x_1x_4+8x_1x_6-2x_2x_3\\
-2x_2x_4+2x_2x_5+2x_2x_6-4x_3x_4+2x_3x_5-10x_4x_5-10x_4x_6+6x_5x_6.   
\end{align*}
Thus, if  $x=(x_1,\ldots,x_6)\in \mathbb{Z}^6$, then $q(x)\in 2\mathbb{Z}$. Hence, $G_4\wedge G_4\in 864\mathbb{Z}$. Therefore, since $G_4\wedge G_4>0$, this minimal norm is 864, which is above the tadpole bound. Despite not identifying which residue forms this corresponds to, and hence being unable to verify if it is a Hodge cycle, we know that no matter what any linear combination of residue forms in this manner will be above 864 as this is the absolute shortest length of the Gram matrix and it is above the bound.

Thus we have verified the tadpole conjecture for this example.

\subsection{A diagonal symmetric example in degree 12}

We now turn to a second example which is much closer in spirit to the sextic: the degree~12 generalized Fermat hypersurface
\[
   X_1 \subset \P_{1,1,1,1,4,4},
\]
given by
\begin{align}
    X_1:= x_0^{12}+x_1^{12}+x_2^{12}+x_3^{12}+x_4^{3}+x_5^{3}.
\end{align}
Again the tame condition is satisfied, since each weight divides the degree:
\[
   1\mid 12,\qquad 4\mid 12.
\]
The hypersurface is elliptically fibered.
One checks that $c_2(X_1)$ is even, so integral fluxes are again allowed:
\[
   G_4+\frac{1}{2}c_2(X_1)\in H^4(X_1,\Z).
\]
The relevant Hodge numbers are
\begin{align}
   h^{1,1}(X_1) = 4,\qquad
   h^{2,2}(X_1) = 3276,\qquad
   h^{3,1}(X_1) = 804,
\end{align}
with induced tadpole bound
\[
   \frac{\chi(X_1)}{24} = 204.
\]

As before we consider maximally symmetric forms.
The Hessian matrix associated with this construction is 2-dimensional, and the symmetric $H^{2,2}$ subspace is also 2-dimensional. This is strikingly close to the sextic situation, where the construction has a very small number of complex-structure moduli and a small symmetric flux lattice.

Concretely, the invariant $H^{2,2}$ forms can be described by the degree-24 monomials with exponent vectors
\[
    \beta = (\beta_0,\dots,\beta_5)
\]
listed here:
\begin{align*}
 [4, 4, 4, 4, 1, 1],\quad
 [6, 6, 6, 6, 0, 0]
\end{align*}
which we interpret as a basis $\{\omega_i\}$ of the symmetric subspace of $H^{2,2}$. Similarly, the invariant $H^{3,1}$ forms are represented by degree-12 monomials with exponent vectors
\begin{align*}
[1, 1, 1, 1, 1, 1],\quad
 [3, 3, 3, 3, 0, 0]
\end{align*}

The Hessian matrix H is given by:
\[
\left(\begin{array}{rr}
\omega_0 & \omega_1 \\
\omega_1 & 0
\end{array}\right)
\]

In that case, the Hessian matrix is much closer to what the sextic example was, since only one pair of $H^{2,2}$ form needs to be turned on to give a Hodge cycle.

The Gram matrix we found is the following : 
\begin{align}
\left(\begin{array}{rrrr}
3456 & -1728 & -1728 & -4320 \\
-1728 & 3456 & 6048 & -1728 \\
-1728 & 6048 & 12096 & -6048 \\
-4320 & -1728 & -6048 & 12096
\end{array}\right)=864\left(\begin{array}{rrrr}
4 & -2 & -2 & -5 \\
-2 & 4 & 7 & -2 \\
-2 & 7 & 14 & -7 \\
-5 & -2 & -7 & 14
\end{array}\right).
\end{align}

It has a signature (4,0) and so is positive definite. This is because the 2 invariant $H^{2,2}$ have Galois action sending them to $H^{4,0}$ and $H^{2,2}$ as well as complex conjugation. Moreover, the quadratic form associated with the matrix on the right side is
\begin{align*}
    q(x)=4x_1^2+4x_2^2+14x_3^2+14x_4^2-4x_1x_2-4x_1x_3-10x_1x_4+14x_2x_3-4x_2x_4-14x_3x_4.
\end{align*}
Thus, if  $x=(x_1,\ldots,x_4)\in \mathbb{Z}^4$, then $q(x)\in 2\mathbb{Z}$. Hence, $G_4\wedge G_4\in 1728\mathbb{Z}$. Therefore, since $G_4\wedge G_4>0$, this minimal norm is 1728, which is above the tadpole bound. 

Its shortest vectors are given by :

norm 1728 : 12 vector(s)
    (2, 0, 1, 1)
    (-2, 0, -1, -1)
    (2, 2, 0, 1)
    (-2, -2, 0, -1)
    (2, 1, 0, 1)
    (-2, -1, 0, -1)
    (1, -1, 1, 1)
    (-1, 1, -1, -1)
    (1, 2, -1, 0)
    (-1, -2, 1, 0)
    (0, 2, -1, 0)
    (0, -2, 1, 0)

norm 3456 : 36 vector(s)
    (4, 1, 1, 2)
    (-4, -1, -1, -2)
    (4, 3, 0, 2)
    (-4, -3, 0, -2)
    (3, -1, 2, 2)
    (-3, 1, -2, -2)
    (3, 1, 1, 2)
    (-3, -1, -1, -2)
    (3, 2, 0, 1)
    (-3, -2, 0, -1)
    (3, 4, -1, 1)
    (-3, -4, 1, -1)
    (1, -2, 2, 1)
    (-1, 2, -2, -1)
    (1, 0, 1, 1)
    (-1, 0, -1, -1)
    (1, 1, 0, 0)
    (-1, -1, 0, 0)
    (1, 3, -1, 0)
    (-1, -3, 1, 0)
    (0, 1, -1, 0)
    (0, -1, 1, 0)
    (0, 1, 0, 0)
    (0, -1, 0, 0)
    (2, -1, 1, 1)
    (-2, 1, -1, -1)
    (2, 3, -1, 1)
    (-2, -3, 1, -1)
    (1, -3, 2, 1)
    (-1, 3, -2, -1)
    (1, 1, 0, 1)
    (-1, -1, 0, -1)
    (1, 0, 0, 0)
    (-1, 0, 0, 0)
    (1, 4, -2, 0)
    (-1, -4, 2, 0)

In this degree-12 example the symmetric lattice is positive definite and the shortest general Hodge cycles have
\[
   L_{\min}=1728,
   \qquad
   \frac{\chi(X_1)}{24}=204 .
\]
Thus,
\[
   \frac{L_{\min}}{\chi(X_1)/24}\simeq 8.4 .
\]
Consequently, within the maximally symmetric sector computed here, every general Hodge cycle overshoots the tadpole bound.  This is an example in which the non-uniform Galois completion already makes the shortest symmetric general Hodge cycles too costly to define supersymmetric Minkowski flux vacua.

\subsection{A degree-8 comparison example}

As a comparison we also consider the degree-8 hypersurface
\[
   X_2\subset \mathbb P_{1,1,1,1,2,2},
   \qquad
   X_2:\; x_0^8+x_1^8+x_2^8+x_3^8+x_4^4+x_5^4=0 .
\]
This example has \(h^{1,1}=1\), but it is still weighted and therefore displays a non-trivial Galois action on the period data.  It provides a useful comparison with both the ordinary sextic and the degree-12 example above.

Restricting to the maximally symmetric \((2,2)\)-sector, the shortest general Hodge cycles have norm
\[
   L_{\min}=512,
   \qquad
   \frac{\chi(X_2)}{24}=113 .
\]
Thus
\[
   \frac{L_{\min}}{\chi(X_2)/24}\simeq 4.5 ,
\]
so the maximally symmetric general Hodge cycles again lie above the tadpole bound.

If one relaxes the symmetry group, shorter vectors appear.  In the examples computed below, these shorter vectors are associated with uniform Galois orbits and do not give the required general Hodge cycles in the maximally symmetric sector.  This supports the interpretation that non-uniform Galois completion is the mechanism responsible for the large tadpole charge of the general Hodge cycles.

 The Gram matrix is given by:
 $$
\left(\begin{array}{rrrr}
2048 & -1024 & -1024 & -512 \\
-1024 & 1024 & 1536 & -512 \\
-1024 & 1536 & 3072 & -1536 \\
-512 & -512 & -1536 & 1536
\end{array}\right)
$$
The few shortest intersections are given by

norm 512 : 8 vector(s)
    (0, 1, -1, -1)
    (0, -1, 1, 1)
    (1, 0, 1, 1)
    (-1, 0, -1, -1)
    (1, 2, 0, 1)
    (-1, -2, 0, -1)
    (1, 1, 0, 1)
    (-1, -1, 0, -1)

norm 1024 : 24 vector(s)
    (1, 1, 0, 0)
    (-1, -1, 0, 0)
    (1, 3, -1, 0)
    (-1, -3, 1, 0)
    (1, 2, -1, 0)
    (-1, -2, 1, 0)
    (1, 0, 1, 2)
    (-1, 0, -1, -2)
    (1, 1, 1, 2)
    (-1, -1, -1, -2)
    (1, -1, 2, 2)
    (-1, 1, -2, -2)
    (2, 2, 1, 2)
    (-2, -2, -1, -2)
    (2, 1, 1, 2)
    (-2, -1, -1, -2)
    (0, 1, -1, 0)
    (0, -1, 1, 0)
    (0, 2, -1, 0)
    (0, -2, 1, 0)
    (2, 3, 0, 2)
    (-2, -3, 0, -2)
    (0, 1, 0, 0)
    (0, -1, 0, 0)

Notably, norm 512 is far above the tadpole already with $\frac{L_{\min}}{\chi/24} \;\gtrsim 4.5$.

Furthermore, taking the next biggest symmetry group, one finds other $h^{2,2}$ forms:
 [[2, 2, 2, 2, 2,
2], [3, 3, 3, 3, 0, 2], [3, 3, 3, 3, 1, 1], [3, 3, 3, 3, 2, 0], [4, 4, 4, 4, 0, 0]]

with shortest vectors : 
norm 96 : 8 vector(s)
    (0, 0, 1, -1, -1, 0, 0)
    (0, 0, -1, 1, 1, 0, 0)
    (1, 1, 1, 0, -1, 0, 0)
    (-1, -1, -1, 0, 1, 0, 0)
    (1, 0, 0, 1, 0, 0, 0)
    (-1, 0, 0, -1, 0, 0, 0)
    (0, 1, 0, 0, 0, 0, 0)
    (0, -1, 0, 0, 0, 0, 0)

norm 128 : 6 vector(s)
    (0, 1, 1, -1, -1, 0, 0)
    (0, -1, -1, 1, 1, 0, 0)
    (1, 0, 1, 0, -1, 0, 0)
    (-1, 0, -1, 0, 1, 0, 0)
    (1, 1, 0, 1, 0, 0, 0)
    (-1, -1, 0, -1, 0, 0, 0)

norm 256 : 12 vector(s)
    (1, 1, 2, -1, -2, 0, 0)
    (-1, -1, -2, 1, 2, 0, 0)
    (1, 0, -1, 2, 1, 0, 0)
    (-1, 0, 1, -2, -1, 0, 0)
    (0, 1, -1, 1, 1, 0, 0)
    (0, -1, 1, -1, -1, 0, 0)
    (1, 2, 1, 0, -1, 0, 0)
    (-1, -2, -1, 0, 1, 0, 0)
    (2, 1, 1, 1, -1, 0, 0)
    (-2, -1, -1, -1, 1, 0, 0)
    (1, -1, 0, 1, 0, 0, 0)
    (-1, 1, 0, -1, 0, 0, 0)

norm 352 : 24 vector(s)
    (0, 1, 2, -2, -2, 0, 0)
    (0, -1, -2, 2, 2, 0, 0)
    (1, 0, 2, -1, -2, 0, 0)
    (-1, 0, -2, 1, 2, 0, 0)
    (1, 2, 2, -1, -2, 0, 0)
    (-1, -2, -2, 1, 2, 0, 0)
    (2, 1, 2, 0, -2, 0, 0)
    (-2, -1, -2, 0, 2, 0, 0)
    (1, 1, -1, 2, 1, 0, 0)
    (-1, -1, 1, -2, -1, 0, 0)
    (1, -1, -1, 2, 1, 0, 0)
    (-1, 1, 1, -2, -1, 0, 0)
    (0, 2, 1, -1, -1, 0, 0)
    (0, -2, -1, 1, 1, 0, 0)
    (1, -1, 1, 0, -1, 0, 0)
    (-1, 1, -1, 0, 1, 0, 0)
    (2, 0, 1, 1, -1, 0, 0)
    (-2, 0, -1, -1, 1, 0, 0)
    (2, 2, 1, 1, -1, 0, 0)
    (-2, -2, -1, -1, 1, 0, 0)
    (2, 1, 0, 2, 0, 0, 0)
    (-2, -1, 0, -2, 0, 0, 0)
    (1, 2, 0, 1, 0, 0, 0)
    (-1, -2, 0, -1, 0, 0, 0)

norm 384 : 8 vector(s)
    (0, 0, 2, -2, -2, 0, 0)
    (0, 0, -2, 2, 2, 0, 0)
    (2, 2, 2, 0, -2, 0, 0)
    (-2, -2, -2, 0, 2, 0, 0)
    (2, 0, 0, 2, 0, 0, 0)
    (-2, 0, 0, -2, 0, 0, 0)
    (0, 2, 0, 0, 0, 0, 0)
    (0, -2, 0, 0, 0, 0, 0)

In accordance with our observations, we note that the only relatively short cycles with respect to the tadpole bound are uniform.

Crucially this is where the Galois action comes into play : the shortest cycles all correspond to uniform $h^{2,2}$ forms, while as soon as the action is non-uniform, it becomes above the tadpole bound.

\subsection{Summary of results}

The computations of this section lead to the following conclusions.

\begin{enumerate}
\item Symmetric flux lattices from cyclotomic periods.
For tame Fermat-type hypersurfaces
\[
   X\subset \mathbb P_{v_0,\ldots,v_5},
   \qquad v_i\mid d,
\]
Movasati's period formula gives an explicit cyclotomic period matrix for the primitive middle cohomology.  After expanding in a rational basis of the cyclotomic field and saturating the resulting lattice, we obtain an integral Gram matrix for the computable symmetric self-dual sector.  This generalizes the Griffiths-residue construction used for the ordinary Fermat sextic to the weighted setting.

\item Degree 36.
For
\[
   X_0\subset \mathbb P_{1,1,1,9,12,12},
   \qquad
   \frac{\chi(X_0)}{24}=597,
\]
we find short vectors of norm at least \(864\) in the computable symmetric general primitive Hodge cycles.  This example, while computationally expensive, showcases our algorithm well and verifies the tadpole conjecture.

\item Degree 12.
For
\[
   X_1\subset \mathbb P_{1,1,1,1,4,4},
   \qquad
   \frac{\chi(X_1)}{24}=204,
\]
the shortest computed symmetric general Hodge cycles have norm
\[
   L_{\min}=1728 .
\]
They therefore overshoot the tadpole bound by a factor of approximately \(8.4\).

\item Degree 8.
For
\[
   X_2\subset \mathbb P_{1,1,1,1,2,2},
   \qquad
   \frac{\chi(X_2)}{24}=113,
\]
the maximally symmetric general Hodge cycles have
\[
   L_{\min}=512,
\]
again above the tadpole bound.

\item Galois action.
The controlled examples show a clear separation between short classes coming from uniform Galois orbits and general Hodge cycles whose Galois completion is non-uniform.  In the latter case the integral reconstruction includes additional middle-cohomology components, and the resulting self-dual norm can become large.  Thus the slogan ``symmetric fluxes are short'' must be refined: in weighted Fermat examples, one must also require compatibility with the Galois action on the period field.
\end{enumerate}

\section{Conclusions}

We have studied \(G_4\)-fluxes and general Hodge cycles for tame Fermat-type Calabi--Yau fourfold hypersurfaces in weighted projective space.  The purpose was to extend the explicit Hodge-theoretic methods developed for the ordinary Fermat sextic to examples where the weighted structure and, in particular, the Galois action on the cyclotomic period data become non-trivial.

The main new observation is that the Galois orbit of a symmetric \((2,2)\)-class need not remain uniform.  In weighted examples, Galois conjugates of a \((2,2)\)-character can contribute to other parts of the middle cohomology.  The rational reconstruction of integral self-dual classes then involves a larger lattice than the naive \((2,2)\)-sector, and this can substantially increase the norm relevant for the tadpole bound.

The degree-$36$ example illustrates the computational difficulty of the method:
although the lattice can be constructed, identifying which integral vectors
correspond to general Hodge cycles requires separating the
$H^{4,0} \oplus H^{0,4}$ block from the $(2,2)$ contribution. Using this
identification, we find that the shortest primitive general Hodge cycles have
norm at least $864$, above $\chi/24 = 597$.

In the simpler degree-12 and degree-8 examples, the shortest computed symmetric general Hodge cycles lie above the tadpole bound.  In the degree-12 case the shortest norm is
\[
   L_{\min}=1728,
   \qquad
   \frac{\chi}{24}=204,
\]
while in the degree-8 comparison example
\[
   L_{\min}=512,
   \qquad
   \frac{\chi}{24}=113.
\]
These examples therefore provide explicit evidence that non-uniform Galois completion can make symmetric general Hodge cycles too costly to support supersymmetric Minkowski vacua.

Taken together, the computations suggest that the arithmetic of the Galois action is an essential ingredient in any extension of the tadpole analysis beyond ordinary projective Fermat hypersurfaces.  They also indicate that singular or quotient geometries remain a natural place to search for flux vacua with reduced tadpole cost, since quotienting can reduce self-intersection numbers while preserving part of the Hodge-theoretic stabilization mechanism.

As an outlook for future research, we hope to return to the study of the Fermat sextic fourfold from a different perspective, namely through the framework developed by Movasati \cite{movasati2025leafschemeshodgeloci}. This approach may provide a more systematic way to understand the relevant Hodge-theoretic and arithmetic structures. We also point out a possible relationship with respect to enumerative geometry as explored in \cite{bae2025countingsurfacescalabiyau4folds} and \cite{bae2024countingsurfacescalabiyau4folds}, provided we better understand the tadpole conjecture and find a framework that would give a satisfying explanation as well as a scalable computational tool.

\bibliographystyle{JHEP}
\bibliography{refs.bib}

\end{document}